\documentclass[preprint,floatfix]{revtex4}
\usepackage[dvips]{graphicx}
\usepackage{epsfig}

\usepackage[cp1251]{inputenc}
\usepackage[english]{babel}
\usepackage{color}
\usepackage{bm}

\usepackage[T1]{fontenc}
\usepackage{xcolor}
\usepackage{lmodern}
\usepackage{listings}
\begin{document}

\title{Flashing a look at the stability of the uniform ferroelectric nematic phase}

\author{E.I.Kats}

\affiliation{Landau Institute for Theoretical Physics, RAS, \\ 
142432, Chernogolovka, Moscow region, Russia}

\begin{abstract}
Recent discovery of the ferroelectric nematic phase $N_F$ resurrects a question about stability of the uniform
$N_F$ state with respect to the formation of either standard for solid ferroelectrics domain structure, or often
occurring in liquid crystals space modulation of the polarization vector ${\bf P}$ (and naturally coupled to ${\bf P}$
nematic director ${\bf n}$). In this work within Landau mean-field theory we investigate the linear stability of the
minimal model admitting the conventional paraelectric nematic $N$ and $N_F$ phases. Our minimal model, (besides the standard
terms of the expansion over the ${\bf P}$ and director gradients) includes, also standard for liquid crystals, director flexoelectric
coupling term ($f$) and often overlooked in the literature (although similar by its symmetry to the director flexoelectric coupling)
the flexo-dipolar coupling ($\beta $). We find that in the easy-plane anisotropy case (when the configuration with ${\bf P}$
orthogonal to ${\bf n}$ is energetically favorable) the uniform $N_F$ state loses its stability with respect to one-dimensional ($1D$) or
two-dimensional ($2D$) modulation. If $f \neq 0$ the $2D$ modulation threshold ($\beta _{c2} $ value) is always higher than its $1D$ 
counterpart value $\beta _{c1}$. No any instability at all if one neglects the flexo-dipolar coupling ($\beta = 0$). In the 
easy-axis case (when ${\bf n}$ prefers to align along ${\bf P}$) the both instability ($1D$ and $2D$) thresholds are the same, 
and the instability
can occur even at $\beta = 0$. We speculate that the phases with $1D$ or $2D$ modulations can be identified with discussed
in the literature [see M.P.Rosseto, J.V.Selinger, Physical Review E, ${\bf 101}$, 052707 (2020)] single splay or double splay nematics.

%\pacs{?????}
\end{abstract}

\date{}

\maketitle

\section{Introduction.}
\label{intro}

Nematics ($N$) are one of the most useful and well-studied class of liquid crystals (see e.g., \cite{GP93}-\cite{BL11}). 
It was a common belief (until relatively 
recent time) that nematics are the only existing in nature state of the achiral matter with unbroken $3D$ translation symmetry,
and partially broken $3D$ rotational symmetry. The revolutionary discovery (in 2011) of so-called twist-bend nematics $N_{TB}$
\cite{CD11} opened a ''Pandora box'' with a plethora of new modulated in space nematic-like phases. To name a few we mention
here well already identified splay nematic phases ($N_S$) \cite{MC18} - \cite{CM20}, and still debated in the literature polar-twisted
nematics (cf., e.g., \cite{CP20} and \cite{DL20}). Eventually along this way of 	chemical synthesis and identification of new nematic-like 
phases, in 2020 the ferroelectric nematic phase $N_F$ was observed in the compound RM734, formed by highly polarizable molecules
(see \cite{CK20}, specially commented in \cite{LA20}). As usual, the discovery of new phases and the elucidation of there
structures is an important topic in physics attracting a lot of attention. It is worth to noting to this $N_F$ discovery, that results
of investigations of the same material (published a bit earlier in \cite{SC20} suggest that the low-temperature state of
the system is formed by alternating domains of the splay nematic phase $N_S$.

All aforesaid developments and results make actual to revisit the stability of the state with uniform dipole polarization in uniaxial
nematics. This is the motivation and the topic of the manuscript. The reminder of it is divided into three sections. In the next
section \ref{basic} we present Landau-like energy expansion for an uniaxial nematic with emerging ferroelectric order. Then
in section \ref{min} we perform linear stability analysis for the set of Euler-Lagrange equations corresponding to the formulated
in section \ref{basic} minimal Landau model. In conclusion section \ref{con} we summarize the results of the work and discuss 
their physical meaning and perspectives for further uses.

\section{Minimal Landau model for ferroelectric nematic ordering.} 
\label{basic}

To formulate our minimal model describing the phase transition into $N_F$ state we start with the ferroelectric part of the 
system free energy. This part is determined by the Landau free energy expansion over polar vector ${\bf P}$ which is
ferroelectric dipole polarization. In a fully isotropic centrosymmetric system the standard form for the average free energy
density reads as
\begin{eqnarray}
\label{pk1}
F^\prime = V^{-1}\int dV \left [\frac{t}{2} {\bf P}^2 + \frac{\lambda }{4}{\bf P}^4 + \frac{b}{2}\left (\nabla {\bf P}\right )^2\right ]
\, ,
\end{eqnarray}
where $V$ is the system volume, and as always in the Landau approach, $t \equiv (T-T_c)/T$ in the vicinity of this (i.e., described
by (\ref{pk1})) phase transition critical temperature $T_c$, and $\lambda $ and $b$ coefficients are supposed to be temperature
($T$) independent. 

However, as it was first noticed by Aslanyan and Levanyuk \cite{AL78} (see also more details in
\cite{BH85}) for the polar vector order parameter ${\bf P}$ even with respect to time inversion (contrary to its pseudo-vector
magnetic counterpart) there is always allowed the non-uniform over ${\bf P}$ contribution
\begin{eqnarray}
\label{pk2}
F^{\prime  \prime }= V^{-1}\int dV \left [\beta _1 {\bf P}^2 \div {\bf P} + \beta _2({\bf P} \nabla ){\bf P}^2\right ]
\, ,
\end{eqnarray}
where $\beta _1$ and $\beta _2$ phenomenological coefficients considered as $T$ independent. 
By its physical meaning the $F^{\prime \prime }$ expansion terms are analogous (under replacement ${\bf n} \to {\bf P}$ to well known in the realm of liquid crystals
flexoelectric energy \cite{GP93}, \cite{KL06}, \cite{BL11}.  In what follows we term the free energy part (\ref{pk2}) by flexo-dipolar
interaction.
Thus the full ferroelectric
part of the free energy expansion is
\begin{eqnarray}
\label{pk3}
F_{P} = F^\prime  + F^{\prime \prime }
\, .
\end{eqnarray}

Since we are interested in the transition into the state with ferroelectric polarization from
the uniaxial nematic phase ($N$), the average free energy density $F_P$ has to be supplemented by the nematic orientation
elasticity (Frank) energy
$F_F$, and the director ${\bf n}$ polarization ${\bf P}$ coupling term. For the sake of simplicity we
assume one elastic constant ($K_F$) approximation for the nematic energy
\begin{eqnarray}
\label{pk4}
F_{F} = \frac{1}{V}\int dV \frac{K_F}{2}\left (\nabla {\bf n}\right )^2
\, ,
\end{eqnarray}
and the simplest form for the coupling term with a single phenomenological coefficient $\gamma_{int}$
\begin{eqnarray}
\label{pk5}
\frac{1}{V}\int dV \gamma _{int}\left ({\bf P}{\bf n}\right )^2
\, .
\end{eqnarray}
On the equal footing with the flexo-dipolar interaction (\ref{pk2}) there is also non-uniform conventional flexoelectric
terms,\cite{GP93}, \cite{KL06}, \cite{BL11},  which couple ${\bf P}$ with the director gradients. Thus we end up with
the following interaction energy
\begin{eqnarray}
\label{pk6}
F_{int} = \frac{1}{V}\int dV \left \{ \gamma _{int}\left ({\bf P}{\bf n}\right )^2 + f_1 {\bf P} {\bf n}div {\bf n} +
f_2 {\bf P}({\bf n}\nabla ){\bf n}\right \}
\, ,
\end{eqnarray}
where $f_1$ and $f_2$ are two nematic flexoelectric coefficients.

For the minimal model, depending on the sign of $\gamma _{int}$ one should distinguish two different initial
configurations: easy-plane case, when $\gamma_{int} > 0$ and easy-axis case for $\gamma _{int} < 0$. In the easy plane case for the
uniformly polarized nematic state, the interaction energy favors to the configuration with ${\bf P}_0$ is orthogonal to ${\bf n}_0$
(where ${\bf P}_0$ is the spontaneous ferroelectric polarization in the uniformly ordered $N_F$ phase, and ${\bf n}_0$ is nematic
director in a such phase). In the easy-axis case ($\gamma _{int} < 0$, the energetically preferable configuration is realized
when ${\bf P}_0$ is parallel to ${\bf n}_0$ (or anti-parallel, because of ${\bf n}_0 \to -{\bf n}_0$ symmetry).

Let us consider first the easy-plane case, assuming that the uniform polarization is along the ${\hat x}$ axis,
and the nematic director  ${\bf n}_0$ is aligned along the ${\hat z}$ axis
\begin{eqnarray}
\label{pk7}
{\bf P}_0 \equiv d_u {\hat x}\, ;\, {\bf n_0} \equiv {\hat z}
\, ,
\end{eqnarray}
where ${\hat x}$ and ${\hat z}$ are corresponding unit vectors.
For such uniform state, the free energy density (\ref{pk1} - (\ref{pk6}) is reduced 
\begin{eqnarray}
\label{pk8}
F_{plane} = \frac{1}{2}t d_u^2 + \frac{1}{4} d_u^4
\, .
\end{eqnarray}
Therefore the magnitude of the spontaneous polarization is
\begin{eqnarray}
\label{pk9}
d_u^2 = -\frac{t}{\lambda }
\, ,
\end{eqnarray}
and the second order transition into this state occurs at $t=0$ independent of $\gamma _{int}$.

In the easy-axis case ($\gamma _{int} <0$) we assume that the both vectors (the uniform polarization  ${\bf P}_0$ and the director
${\bf n}_0$ ) are aligned parallel to the ${\hat x}$ axis. Then, the free energy density (\ref{pk1} - (\ref{pk6}) reads as
\begin{eqnarray}
\label{pk10}
F_{axis} = \frac{1}{2}t d_u^2 + \frac{1}{4} d_u^4 + \gamma _{int}d_u^2
\, .
\end{eqnarray}
The free energy density (\ref{pk10}) tells that the second order phase transition critical temperature is shifted to ${\tilde t} = 0$,
where
\begin{eqnarray}
\label{pk11}
{\tilde t} \equiv t - 2|\gamma _{int}|
\, ,
\end{eqnarray}
and
\begin{eqnarray}
\label{pk12}
d_u^2 = -\frac{{\tilde t}}{\lambda }
\, .
\end{eqnarray}
In the next section \ref{min} we investigate the stability of the uniform ferroelectric state for the easy-plane and easy-axis
signs of $\gamma _{int}$.

\section{Stability analysis of the minimal model.} 
\label{min}
As it was mentioned in the previous section \ref{basic}
the types of the ordering in the easy-plane and easy-axis configurations are different ones. Therefore we analyze the stability
of the corresponding uniformly polarized states separately for $\gamma _{int} >0$ and $\gamma _{int} < 0$.
\subsection{Easy-plane case}
\label{plane}
In the case with $\gamma _{int} > 0$, the unperturbed uniform state (see (\ref{pk7}) is ${\bf P}_0 = d_u {\hat x}$, and ${\bf n}_0 = {\hat z}$.
Non-uniform (space dependent) perturbations (needed to study linear stability) of this state can be defined in the following form
\begin{eqnarray}
\label{pk13}
{\bf P}_0 = (d_u + \epsilon _1){\hat x} + \epsilon _2{\hat y} + \epsilon _3{\hat z} \, ,\, \delta {\bf n} = \delta n_x {\hat x} +
\delta n_y {\hat y}
\, ,
\end{eqnarray}
where $\epsilon _1({\bf r})$ $\epsilon _1({\bf r})$, $\epsilon _2({\bf r})$, $\epsilon _3({\bf r})$, $\delta n_x({\bf r})$,
and $\delta n_y({\bf r})$ are small, space dependent perturbations of the uniform state. Expanding the average free energy density
(\ref{pk1}) - {\ref{pk6}) over these perturbations (up to the quadratic terms one only needs to study the linear stability),
 we get the following perturbation free-energy
\begin{eqnarray}
\label{pk14}
\delta f = \frac{t}{2}\left [2d_u \epsilon _1 + \epsilon _1^2 + \epsilon _2^2 + \epsilon _3^2\right ]
+\frac{\lambda }{4}\left [4d_u^3\epsilon _1 + 6d_u^2\epsilon _1^2 + 2 d_u^2\epsilon _2^2 + 2d_u^2\epsilon _3^2\right ]\\
\nonumber
+ 2\beta d_u\left [\epsilon _1\frac{\partial \epsilon _2}{\partial y} + \epsilon _1\frac{\partial \epsilon _3}{\partial z}\right ]
+ \gamma _{int}\left [d_u^2 (\delta n_x)^2 + 2 d_u \delta n_x \epsilon _3\right ] \\
\nonumber
+ f\left [\epsilon _3\frac{\partial \delta n_x}{\partial x} + \epsilon _3\frac{\partial \delta n_y}{\partial y}\right ]
+ \frac{b}{2}\left (\nabla {\bf P}\right )^2 + \frac{K_F}{2}\left (\nabla {\bf n}\right )^2
\, .
\end{eqnarray}
It is worth to noting that the both flexo-dipolar terms (with the coefficients $\beta _1$ and $\beta _2$) can be transformed one into another by adding the terms with full space derivatives. Similarly the both flexoelectric terms  in our approximation can be replaced by the single term.
For these reasons (and to get a bit simpler equations) we neglect the surface energy contributions and left in
(\ref{pk14}) only one flexo-dipolar term and as well only one flexoelectric term.

With this expression in hands, we can derive the Euler-Lagrange equations for the perturbations, and the stability with respect to these 
perturbations of the uniformly
polarized $N_F$ state is the condition that the Euler-Lagrange equations have non-zero solutions. We omit these very simple and standard calculations and write down the final stability condition in the Fourier representation ($\epsilon _1\, ,\, \epsilon _2\, ,\, \epsilon _3\,
,\, \delta n_x\, ,\, \delta n_y\, \propto \exp (i{\bf q}{\bf r})$
\begin{eqnarray}
\label{pk15}
(2 + k^2)\left [\sigma _1 - \sigma _2\right ] = \frac{4\beta ^2}{b \lambda }\left [\sigma _1(\alpha _y^2 + \alpha _z^2) - 
\sigma _2\alpha _y^2 \right ]
\, .
\end{eqnarray}
Here we utilize the dimensionless wave vectors (in units of the correlation radius $r_c = (b/|t|)^{1/2}$) ${\bf k} \equiv {\bf q} r_c$,
and $\alpha _i = k_i/k$ (where $i =x , y , z$). Besides for the sake of compactness the following notations are introduced in (\ref{pk15})
\begin{eqnarray}
\label{pk16}
\sigma _1 = \frac{K_F}{b} k^2\left (\frac{2\gamma _{int}}{\lambda } + \frac{K_F}{b} k^2\right )\\
\nonumber
\sigma _2 = \frac{K_F}{b}\left (\frac{4 \gamma _{int}^2}{\lambda |t|} + \frac{f^2}{b |t|}\alpha _x^2\right ) +
\frac{f^2}{b|t|}\alpha ^2\left (\frac{2\gamma _{int}}{\lambda } + \frac{K_F}{b} k^2\right ) 
\, .
\end{eqnarray}
The above stability criterion (\ref{pk16}) allows us to conclude
\begin{itemize}
\item
For the easy-plane case, there is no any instability at all without flexo-dipolar interaction (i.e., at $\beta =0$).
\item
If $\beta \neq 0$ and the standard flexoelectric coefficient $f \neq 0$, the double-splay instability occurs always
at higher $\beta $-values than the single-splay instability. The position of the threshold (i.e., the critical value of $\beta $) depends on the system parameters
($f , k_F , b , \gamma _{int}$) and temperature $t$.
\end{itemize}

\subsection{Easy-axis case}
\label{axis}

For the easy-axis case ($\gamma _{int} < 0$) the generic form of the perturbation (relative to the
initial configuration with ${\bf P}_0 = d_u {\hat x}$ and ${\bf n}_0 = {\hat x}$) can be represented as
\begin{eqnarray}
\label{pk17}
{\bf P} = (d_u + \epsilon _1){\hat x} + \epsilon _2{\hat y} + \epsilon _3{\hat z} \, ,\, \delta {\bf n} = \delta n_z {\hat z} +
\delta n_y {\hat y}
\, .
\end{eqnarray}
Then the perturbative part of the free energy density reads as
\begin{eqnarray}
\label{pk18}
\delta f = \frac{t}{2}\left [2d_u \epsilon _1 + \epsilon _1^2 + \epsilon _2^2 + \epsilon _3^2\right ]
+\frac{\lambda }{4}\left [4d_u^3\epsilon _1 + 6d_u^2\epsilon _1^2 + 2 d_u^2\epsilon _2^2 + 2d_u^2\epsilon _3^2\right ]\\
\nonumber
+ 2\beta d_u\left [\epsilon _1\frac{\partial \epsilon _2}{\partial y} + \epsilon _1\frac{\partial \epsilon _3}{\partial z}\right ]
- |\gamma _{int}|\left [2 d_u \epsilon _1 + \epsilon _1^2 + 2 d_u \epsilon _2 \delta n_y) + 2 d_u \delta n_z \epsilon _3\right ] \\
\nonumber
+ f \epsilon _1 \left [\frac{\partial \delta n_y}{\partial y} + \frac{\partial \delta n_z}{\partial z}\right ]
+ \frac{b}{2}\left (\nabla {\bf P}\right )^2 + \frac{K_F}{2}\left (\nabla {\bf n}\right )^2
\, ,
\end{eqnarray}
where for the easy-axis case $d_u^2 = |{\tilde t}|/\lambda $ with ${\tilde t} \equiv t - 2|\gamma _{int}|$, and it is convenient
to use $r_c^2 = b/|{\tilde t}|$ to define the dimensionless wave-vector $k$.

Minimizing the free-energy density (\ref{pk18}) over the perturbations (\ref{pk17}) we arrive at the linearized Euler-Lagrange equations.
Then the stability of the uniform $N_F$ phase is determined by the following condition
\begin{eqnarray}
\label{pk19}
(2 + k^2)\Sigma = \left [\frac{4\beta ^2 K_F}{b \lambda } k^4 + \frac{8\beta f |\gamma _{int}}{\lambda b|{\tilde t|}} +
\frac{f^2}{b |{\tilde t}|}\left (k^2 + \frac{2|\gamma _{int}|}{|{\tilde t}|}\right )\right ]\left (\alpha _y^2 + \alpha _z^2\right )
\, ,
\end{eqnarray}
where
\begin{eqnarray}
\label{pk20}
\Sigma =  \frac{K_F k^2}{b}\left (k^2 + \frac{2|\gamma _{int}|}{|{\tilde t}|}\right ) - \frac{4\gamma _{int}^2}{\lambda |{\tilde t}|}
\, .
\end{eqnarray}
By a simple inspection of the expressions (\ref{pk19}) - (\ref{pk20}) we arrive eventually at the following conclusions
\begin{itemize}
\item
For the easy-axis configuration, the stability threshold of the uniform $N_F$ state
is degenerate for $1D$ and $2D$ modulations (we identify with a single-splay and double-splay phases respectively).
\item
The threshold position depends on all model parameters ($\beta  , f , k_F , b , \gamma _{int}$) and the shifted temperature ${\tilde t}$.
\item
In the case of non-zero flexoelectric coupling (i.e., if $f\neq 0$) the instability may occur even without the flexo-dipolar interaction ($\beta = 0$).
\end{itemize}
These two (easy-plane and easy-axis) stability conditions (\ref{pk19}) - (\ref{pk20}) and (\ref{pk15}) - (\ref{pk16}) are our main results
in this work.

\section{Conclusion and Perspectives.}
\label{con}
Recent discoveries of new types of liquid crystals (ferroelectric nematics $N_F$ \cite{CK20}, splay-nematics \cite{SC20},
twist-bend nematics \cite{CD11} and polar-twisted nematics (\cite{CP20}, \cite{DL20}) set challenges to look to
the macroscopic phase behavior of the new phases within a simple model . We do believe that the proposed in this work minimal model
is indeed as simple as it is possible to keep all essential features of the phase diagrams. The model includes the relevant
nematic and ferroelectric degrees of freedom and their uniform ($\gamma _{int}$) and non-uniform (flexo-dipolar ($\beta $) and flexoelectric
($f$) couplings. Performed in the work theoretical analysis of the minimal model predicts that the uniformly polarized $N_F$ phase can become unstable with respect to the space modulations of the polarization (${\bf P}$) (and coupled to ${\bf P}$ nematic director ${\bf n}$). The instability occurs due to the non-uniform coupling terms (flexo-dipolar $\beta $, and flexoelectric $f$) and only when the coupling strengths
are higher than certain threshold values. For the easy-plane case, there is no any instability at all without flexo-dipolar interaction (i.e., at $\beta =0$). If $\beta \neq 0$ and the standard flexoelectric coefficient $f \neq 0$, the double-splay instability occurs always
at higher $\beta $-values than the single-splay instability. The position of the threshold (i.e., the critical value of $\beta $) depends on the system parameters
($f , k_F , b , \gamma _{int}$) and temperature $t$. For the easy-axis configuration, the stability threshold of the uniform $N_F$ state
is degenerate for $1D$ and $2D$ modulations (we identify with a single-splay and double-splay phases respectively).
To remove the degeneracy the higher order terms over the perturbations should be included in the free-energy expansion.
The threshold position depends on all model parameters ($\beta  , f , k_F , b , \gamma _{int}$) and the shifted by the uniform coupling $\gamma _{int}$ temperature ${\tilde t} = t - 2|\gamma _{int}|$.
In the case of non-zero flexoelectric coupling (i.e., if $f\neq 0$) the instability may occur even without the flexo-dipolar interaction ($\beta = 0$).

Our simple model neglects some elements (e.g., non-linear higher order terms in the Landau energy, or thermal fluctuations)
which can modify quantitatively (and even qualitatively) the predicted phase
diagram. Calculations with such elements taken into consideration as realistically as possible is 
however doomed to be prohibitively  bulky. Such more elaborated and specific study might become
appropriate should suitable experimental 
results become available. It is not the case for the moment, and transparency of our consideration is worth a few oversimplifications.
Besides performed in the work  simple
calculations are nevertheless instructive.

\vspace{0.5cm}

\acknowledgments
I am grateful to V.Lebedev, B.Ostrovskii, and E.Pikina for stimulating discussions and helpful comments

\end{document}